\documentclass[fdp,a4paper]{w-art}
\usepackage{times,cite,w-thm}
\usepackage{epsfig}

\usepackage{bbm}
\usepackage{amssymb,amsmath}

\newcommand{\be}{\begin{equation}}
\newcommand{\ee}{\end{equation}}
\newcommand{\ba}{\begin{eqnarray}}
\newcommand{\ea}{\end{eqnarray}}
\newcommand{\no}{\nonumber\\}

\newcommand{\mnu}{\mathcal{M}_\nu}

\newcommand{\zz}{\mathbbm{Z}_2}

\begin{document}

\title{$\mu$--$\tau$ interchange symmetry and lepton mixing}

\author[W. Grimus]{Walter Grimus\inst{1,}%
\footnote{Corresponding author\quad 
E-mail:~\textsf{walter.grimus@univie.ac.at},
Phone: +43-4277-51519
Fax: +43-4277-9515}}
\address[\inst{1}]{University of Vienna, Faculty of Physics, 
Boltzmanngasse 5, A--1090 Vienna, Austria}
\author[L. Lavoura]{Lu\'{\i}s Lavoura\inst{2,}%
\footnote{E-mail:~\textsf{balio@cftp.ist.utl.pt}}}
\address[\inst{2}]{Technical University of Lisbon
and CFTP, 
Instituto Superior T\'ecnico, 1049-001 Lisbon, Portugal}

\begin{abstract}
We focus on the use of a $\mu$--$\tau$ interchange symmetry
to explain features of lepton mixing,
especially maximal atmospheric neutrino mixing.
We review two models which achieve this goal
and are based on the seesaw mechanism
and on the soft breaking of the 
family-lepton-number symmetries.
We also note that that symmetry may be embedded
in a generalized CP symmetry.
We show that,
in the context of some of our models,
arguments of naturalness may be used
for explaining the smallness of the mass ratio $m_\mu / m_\tau$.
\end{abstract}

\maketitle

\section{Introduction}

Experiments on neutrino oscillations
have greatly enlarged our knowledge
on neutrino masses and lepton mixing~\cite{rpp}.
The lepton mixing matrix
is quite different from the quark mixing (CKM) matrix~\cite{rpp},
since it features two large mixing angles---the angles
$\theta_{12}$ and $\theta_{23}$
for the solar and atmospheric neutrino oscillations, respectively---while
the third mixing angle,
$\theta_{13}$,
which is relevant for instance in the oscillations of reactor neutrinos,
is small~\cite{results,schwetz}. 

Much effort has been devoted
for quite some time 
to model building for neutrino masses and lepton mixing,
with focus on possible underlying flavour symmetries;
for recent reviews see,
for instance,
refs.~\cite{review} and~\cite{finitegroups}
for the model-building and group-theoretical,
respectively,
aspects of that problem.
In the basis where the charged-lepton mass matrix is diagonal,
a $\mu$--$\tau$ interchange symmetry~\cite{mu-tau-old}
$\nu_{\mu L} \leftrightarrow \nu_{\tau L}$
in the neutrino mass terms
ensures a form of the (effective) neutrino mass matrix $\mnu$
which simultaneously gives $\theta_{23} = 45^\circ$ and $\theta_{13} = 0$.
A stronger restriction on $\mnu$
leads to tri-bimaximal mixing~\cite{HPS} in which,
in addition to the predictions of $\mu$--$\tau$ symmetry,
one also has $\sin^2 \theta_{12} = 1/3$;
however,
the models which aim at predicting tri-bimaximal mixing are,
in our opinion,
very much contrived.
On the other hand,
as we shall show in this review,
less ambitious models which aim at
including only the $\mu$--$\tau$ symmetry
are relatively simple.

Recently,
evidence of $\theta_{13} \neq 0$ has been found~\cite{schwetz};
this reduces the $\mu$--$\tau$ interchange symmetry
to just an approximate symmetry.
Several strategies have been proposed in this context:
\begin{itemize}
\item Some terms in the Lagrangian may violate explicitly
the $\mu$--$\tau$ symmetry.
This has been discussed for instance in ref.~\cite{mutau-broken}.
We shall not pursue that possibility any further in this review.
\item Radiative corrections to $\mnu$
induced by the renormalization-group running
may disturb the $\mu$--$\tau$ interchange symmetry.
In this mechanism,
the $\mu$--$\tau$ symmetry
is assumed to be exact in $\mnu$ at some high energy scale,
\textit{e.g.}\ at the seesaw scale~\cite{seesaw};
the renormalization-group evolution
of the neutrino mass operators
from that high scale to the electroweak scale
produces a mixing among those operators which,
after insertion of the vacuum expectation values (VEVs),
adds a $\mu$--$\tau$-antisymmetric contribution
to the original $\mu$--$\tau$-symmetric part of $\mnu$~\cite{GLrenorm}.
Note that this mechanism must be active in any realistic model,
since scalars odd under the $\mu$--$\tau$ interchange symmetry
are necessarily present in order to achieve $m_\mu \neq m_\tau$.
As noted in ref.~\cite{GLrenorm},
however,
this mechanism requires a quasi-degenerate neutrino mass spectrum
in order to produce a sizeable $\theta_{13}$;
such a quasi-degenerate spectrum is on the verge of being ruled out
by cosmological arguments~\cite{rpp}.
\item One may replace the $\mu$--$\tau$ symmetry
by a generalized CP symmetry which includes $\mu$--$\tau$ interchange.
This mechanism~\cite{HScp,GLcp},
which will be discussed in this review,
leaves $\theta_{13}$ free and predicts instead maximal CP violation
via the CKM-type phase $\delta$ of the lepton mixing matrix.
\end{itemize}

In the construction of $\mu$--$\tau$-symmetric models,
the most straightforward mechanism for achieving small neutrino masses
is the seesaw mechanism~\cite{seesaw}.
In a minimal version of that mechanism one may content oneself
with two right-handed neutrino gauge singlets,
as was done for instance in refs.~\cite{baba,he}.
In this review,
however, 
we shall always use three heavy neutrino singlets,
and also assume the existence of only three light neutrinos.

For completeness we want to mention here some aspects
which have been investigated
in the framework of the $\mu$--$\tau$ interchange symmetry
but will not be discussed in this review:
\begin{itemize}
\item The $\mu$--$\tau$ symmetry may be incorporated 
in grand unified models based on the gauge group $SU(5)$~\cite{GLsu5,mohapatra}.
\item The $\mu$--$\tau$ symmetry may be extended to the quark sector,
thus promoting it to a universal flavour symmetry~\cite{universal}.
\item Leptogenesis can successfully be implemented
in $\mu$--$\tau$-symmetric models~\cite{leptogenesis}.
\item
The $\mu$--$\tau$ symmetry may be supplemented by further conditions,
for instance by texture zeros, in order 
to enhance the predictivity of a model~\cite{roy}.
\end{itemize}

This review is organized as follows.
After section~2,
which contains some elementary notation,
we proceed in section~3 to expose the consequences for lepton mixing
of a $\mu$--$\tau$-symmetric neutrino mass matrix $\mnu$,
and of a $\mnu$-symmetric under a generalized CP transformation
involving the $\mu$--$\tau$ interchange.
Sections~4 and~5 contain specific models
which lead to the neutrino mass matrices studied in section~3.
Section~6 is devoted to the addition,
to the models of sections~4 and~5,
of an extra symmetry $K$
which allows one to explain the smallness of the ratio $m_\mu / m_\tau$
as the natural result of the small soft breaking of $K$.
Section~7 shortly summarizes the contents of this review.

\section{Notation}

Let $\mnu$ denote the light-neutrino (effective) Majorana mass matrix
in the weak basis where the charged-lepton mass matrix is diagonal.
Thus,
\ba
\mathcal{L}_\mathrm{lepton \ mass} &=&
-
\left( \begin{array}{ccc} \bar e_L, & \bar \mu_L, & \bar \tau_L
\end{array} \right)
\left( \begin{array}{ccc} m_e & 0 & 0 \\ 0 & m_\mu & 0 \\ 0 & 0 & m_\tau
\end{array} \right)
\left( \begin{array}{c} e_R \\ \mu_R \\ \tau_R \end{array} \right)
\no & &
+ \frac{1}{2}
\left( \begin{array}{ccc} \nu_{eL}^T, & \nu_{\mu L}^T, & \nu_{\tau L}^T
\end{array} \right) \mnu C^{-1}
\left( \begin{array}{c} \nu_{eL} \\ \nu_{\mu L} \\ \nu_{\tau L}
\end{array} \right) + \mathrm{H.c.},
\label{svirt}
\ea
where $C$ is the Dirac--Pauli charge-conjugation matrix.
The matrix $\mnu$ is symmetric,
hence it can be bi-diagonalized by a single unitary matrix $U$ via 
\be
U^T \mnu U = \mathrm{diag} \left( m_1, \ m_2, \ m_3 \right) 
\equiv \hat m,
\label{diag}
\ee
where the $m_j$ ($j = 1, 2, 3$) are non-negative real.
Equation~(\ref{diag}) means that
\be
\mnu c_j = m_j c_j^\ast
\label{eigen}
\ee
(no sum over $j$ is assumed),
where the $c_j$ denote the three columns
of $U = \left( c_1, \ c_2, \ c_3 \right)$.

If we consider the charged-current Lagrangian
\ba
\mathcal{L}_{\mathrm{lepton}\mbox{--}W} &=&
\frac{g}{\sqrt{2}}\, W^-_\mu
\left( \begin{array}{ccc} \bar e_L, & \bar \mu_L, & \bar \tau_L
\end{array} \right) \gamma^\mu
\left( \begin{array}{c} \nu_{eL} \\ \nu_{\mu L} \\ \nu_{\tau L}
\end{array} \right) + \mathrm{H.c.}
\no
&=& \frac{g}{\sqrt{2}}\, W^-_\mu
\left( \begin{array}{ccc} \bar e_L, & \bar \mu_L, & \bar \tau_L
\end{array} \right) \gamma^\mu U
\left( \begin{array}{c} \nu_1 \\ \nu_2 \\ \nu_3
\end{array} \right) + \mathrm{H.c.},
\label{hbvue}
\ea
we see that $U$ is,
indeed,
the lepton mixing (PMNS) matrix.

\section{$\mu$--$\tau$ interchange symmetry}

A $\mu$--$\tau$-symmetric $\mnu$ is,
by definition,
of the form
\be
\mnu = \left( \begin{array}{ccc} x & y & y \\ y & z & w \\ y & w & z
\end{array} \right),
\label{rtyde}
\ee
where $x$,
$y$,
$z$,
and $w$ are in general complex.
The mass matrix $\mnu$ of equation~(\ref{rtyde}) can be characterized 
as being the general solution of the equation~\cite{GLmaxmix}
\be\label{SMS+}
S \mnu S = \mnu
\ee
with
\be\label{S}
S = \left( \begin{array}{ccc} 1 & 0 & 0 \\ 0 & 0 & 1 \\ 0 & 1 & 0
\end{array} \right).
\ee
Notice that the Lagrangian
can never enjoy full $\mu$--$\tau$ interchange symmetry,
since the masses of the muon ($m_\mu$) and tau ($m_\tau$) charged leptons
differ ($m_\mu \neq m_\tau$);
thus,
a $\mu$--$\tau$ interchange symmetry exists,
at most,
in the neutrino mass matrix.

Since the $\mu$--$\tau$ interchange symmetry,
\textit{viz.}\ the matrix $S$ of equation~(\ref{S}),
is a $\zz$ symmetry,
it has eigenvalues $\pm 1$.
The matrix in equation~(\ref{rtyde}) corresponds to the eigenvalue
$+1$,
\textit{cf.}\ equation~(\ref{SMS+}).
Corresponding to the eigenvalue $-1$,
one might consider~\cite{mutau-anti,GKLST} a $\mu$--$\tau$-antisymmetric
\be
\mnu = \left( \begin{array}{ccc} 0 & y & -y \\ y & z & 0 \\ -y & 0 & -z
\end{array} \right),
\label{jvirt}
\ee
which is the solution of $S \mnu S = -\mnu$.
However,
the matrix in equation~(\ref{jvirt}) leads
not only to maximal atmospheric mixing
but also to maximal solar mixing
and to two degenerate neutrinos~\cite{GKLST} 
and can therefore,
at best,
serve as a first approximation for the true $\mnu$.
We shall not consider equation~(\ref{jvirt}) any further in this review.

Because of the freedom of rephasing the lepton fields
in equations~(\ref{svirt}) and~(\ref{hbvue}),
the matrix $\mnu$ in equation~(\ref{rtyde}) may be generalized
and we need only require
\ba
\left| \left( \mnu \right)_{\mu\mu} \right|
&=&
\left| \left( \mnu \right)_{\tau\tau} \right|,
\no
\left| \left( \mnu \right)_{e\mu} \right|
&=&
\left| \left( \mnu \right)_{e\tau} \right|,
\label{hvuer} \\
\arg{\left[ \left( \mnu \right)_{ee} \left( \mnu \right)_{\mu\mu}
{\left( \mnu \right)_{e\mu}^\ast}^2 \right]}
&=& \arg{\left[ \left( \mnu \right)_{ee} \left( \mnu \right)_{\tau\tau}
{\left( \mnu \right)_{e\tau}^\ast}^2 \right]}
\nonumber
\ea
in order to obtain a $\mu$--$\tau$-invariant $\mnu$.
Equations~(\ref{hvuer}) constitute two constraints on the moduli
and one constraint on a phase of the matrix elements of $\mnu$.
We expect an identical number of constraints
on the lepton-mixing observables
to follow from the three conditions~(\ref{hvuer}).

Indeed,
the characteristic feature of the matrix $\mnu$ in equation~(\ref{rtyde})
is that it has an eigenvector 
$\left( 0, \, 1, \, -1 \right)$
corresponding to the eigenvalue $z - w$.
Let $\varphi = \arg{\left( z - w \right)}$,
then
\be
\mnu
\left( \begin{array}{c} 0 \\ e^{- i \varphi / 2} \\ - e^{- i \varphi / 2}
\end{array} \right) = \left| z - w \right|
\left( \begin{array}{c} 0 \\ e^{i \varphi / 2} \\ - e^{i \varphi / 2}
\end{array} \right).
\label{wvyur}
\ee
This is an instance of equation~(\ref{eigen}).
One thus finds that a $\mu$--$\tau$-symmetric $\mnu$ 
leads to one of the columns of the PMNS matrix being 
$\left( 0, \, e^{- i \varphi / 2}, \, - e^{- i \varphi / 2} \right)^T$.
Phenomenologically,
this is realistic only if that column is the third one,
\textit{i.e.}\ the one corresponding to the neutrino ($\nu_3$)
which has a mass ($m_3$) most afar from the other two.
Indeed,
phenomenologically $U_{e3}$ is very close to zero
but $U_{e1}$ and $U_{e2}$ are certainly non-zero. 
We thus conclude that a $\mu$--$\tau$-symmetric $\mnu$ leads to
\ba
U_{e3} &=& 0,
\label{vjura} \\
\theta_{23} &=& 45^\circ;
\label{iuty}
\ea
the condition~(\ref{iuty}) is equivalent to
$\left| U_{\mu 3} \right| = \left| U_{\tau 3} \right|$,
\textit{viz.}\ maximal atmospheric neutrino mixing.
Since the condition~(\ref{vjura}) entails that the Dirac phase $\delta$
is meaningless and may be removed from $U$,
equations~(\ref{vjura}) and~(\ref{iuty}) in fact constitute
two constraints on the moduli
and one constraint on a phase of the PMNS matrix elements,
as expected.

The matrix $\mnu$ in equation~(\ref{rtyde}) does not impose
any constraint on the neutrino masses.
One has,
from equation~(\ref{wvyur}),
\be
m_3 = \left| z - w \right|.
\ee
The other two masses may be obtained by considering
the determinant and the trace of $\mnu \mnu^\ast$:
\ba
m_1 m_2 &=& \left| x \left( z + w \right) - 2 y^2 \right|,
\\
m_1^2 + m_2^2 &=& \left| x \right|^2 + \left| z + w \right|^2
+ 4 \left| y \right|^2.
\ea
The solar mixing angle is given by~\cite{GLmu-tau}
\be
\tan{2 \theta_{12}} = 
\frac{2 \sqrt{2} \left| x^\ast y + y^\ast \left( z + w \right) \right|}
{\left| z + w \right|^2 - \left| x \right|^2}.
\ee
Note that tri-bimaximal mixing,
\textit{i.e.}\ $\tan{2 \theta_{12}} = 2 \sqrt{2}$,
ensues if some additional symmetry enforces
$\left| x^\ast y + y^\ast \left( z + w \right) \right|
= \left| \left| x \right|^2 - \left| z + w \right|^2 \right|$.
This is the case,
in particular,
if the $\mu$--$\tau$-symmetric $\mnu$
has the additional property that
the sums of its matrix elements over each row are all equal,
\textit{viz.}\ $x + y = z + w$.

Besides $\mu$--$\tau$ interchange,
one may consider the superposition
of the $\mu$--$\tau$ interchange on a CP transformation,
\textit{i.e.}\ one may require invariance of the mass Lagrangian
in the second line of equation~(\ref{svirt})
under the generalized CP transformation~\cite{HScp,GLcp}
\ba
\nu_{eL} &\to& i \gamma^0 C \bar \nu_{eL}^T, \no
\nu_{\mu L} &\to& i \gamma^0 C \bar \nu_{\tau L}^T, \label{cp} \\
\nu_{\tau L} &\to& i \gamma^0 C \bar \nu_{\mu L}^T. \nonumber
\ea
That invariance requirement leads to~\cite{GLcp,GLmaxmix}
\be
\mnu = \left( \begin{array}{ccc} a & r & r^\ast \\ r & s & b \\
r^\ast & b & s^\ast
\end{array} \right),
\label{rtydr}
\ee
where $r$ and $s$ are in general complex but $a$ and $b$ are real.
The $\mnu$ in equation~(\ref{rtydr}) is the general solution of the equation
\be\label{SMS*}
S \mnu S = \mnu^\ast.
\ee
The neutrino mass matrix of equation~(\ref{rtydr})
may also be obtained from an $A_4$ model~\cite{ma}
without invoking invariance under the CP transformation of equation~(\ref{cp}).

What are the predictions the mass matrix of equation~(\ref{rtydr})? 
We first use equation~(\ref{diag}) to express that matrix as 
$\mnu = U^\ast \hat m U^\dagger$.
This allows one to reformulate equation~(\ref{SMS*}) as 
\be\label{X}
X \hat m = \hat m X^\ast,
\ee
with
\be
\label{xdef}
X = U^\dagger S U^\ast.
\ee
The matrix $X$ is symmetric and unitary.
Equation~(\ref{X}) has the following real and imaginary components:
\ba
\left( m_i - m_j \right) \mathrm{Re}\, X_{ij} &=& 0, \label{re} \\ 
\left( m_i + m_j \right) \mathrm{Im}\, X_{ij} &=& 0. \label{im}
\ea
Since both the solar and the atmospheric neutrino mass-squared differences
are non-zero,
$\hat m$ is non-degenerate.
Consequently,
from equation~(\ref{re}),
$\mathrm{Re}\, X_{ij} = 0$ if $i \neq j$.
Moreover,
the neutrino masses $m_j$ are non-negative
and at most one of them may vanish.
Consequently,
from equation~(\ref{im}),
$\mathrm{Im}\, X_{ij} = 0$ if $i \neq j$.
Thus,
$X$ is diagonal.
Since $X$ is unitary,
its diagonal elements are phases.
Because of equation~(\ref{im}),
$X_{jj} = \pm 1$ whenever $m_j$ is non-zero.
If $m_j=0$,
then $X_{jj}$ is an arbitrary phase;
however,
in that case one can absorb a factor $(X_{jj})^{-1/2}$
into the neutrino field $\nu_j$,
whereupon $X_{jj}$ becomes equal to $1$.
Now,
we rewrite equation~(\ref{xdef}) as
\be\label{SUUX}
S U^\ast = U X.
\ee
Equation~(\ref{SUUX}) determines the following forms
for the columns $c_j$ of $U$~\cite{GLcp}:
\be
\label{Xc}
\begin{array}{rcl}
X_{jj} = 1 & \Rightarrow &
{\displaystyle c_j =
\left( \begin{array}{c} u_j \\ w_j \\ w_j^\ast \end{array} \right)},
\\*[3mm]
X_{jj} = -1 & \Rightarrow &
{\displaystyle c_j =
\left( \begin{array}{c} iu_j \\ w_j \\ -w_j^\ast \end{array} \right)},
\end{array}
\ee
with real $u_j$.
One reads off from equations~(\ref{Xc}) that~\cite{HScp,GLcp}
\be
\left| U_{\mu j} \right| = \left| U_{\tau j} \right|
\quad \mbox{for} \ j=1,2,3.
\ee
Inserting this condition into the standard parameterization of $U$
yields the predictions
\ba
\theta_{23} &=& 45^\circ,
\\
\sin{\theta_{13}} \cos{\delta} &=& 0.
\ea
Moreover, equations~(\ref{Xc}) show that
the relative phases among $U_{e1}$,
$U_{e2}$,
and $U_{e3}$ are multiples of $90^\circ$;
this means that the Majorana phases,
which arise for instance in the computation
of the effective mass relevant 
for neutrinoless $\beta\beta$ decay,
have values either zero or $\pi$.

\section{Softly broken lepton numbers and $\mu$--$\tau$ interchange symmetry}
\label{softly broken}

In this section we discuss a scenario
where the $\mu$--$\tau$ interchange symmetry
is embedded in a model
with softly broken family lepton numbers~\cite{GLmu-tau}.
Let $\zz^{(\mu\tau)}$ denote the cyclic group of order two 
generated by the $\mu$--$\tau$ interchange symmetry.
The model of ref.~\cite{GLmu-tau} is a (type~I) seesaw model
in which three right-handed heavy neutrino singlets $\nu_{\alpha R}$
($\alpha = e,\mu,\tau$)
are added to the Standard Model multiplets.
Moreover,
as we will see shortly,
the model needs three Higgs doublets $\phi_j$. 
The symmetries of the model are the following:
\begin{enumerate}
\renewcommand{\labelenumi}{\roman{enumi})}
\item The $U(1)_{L_\alpha}$ associated with the family lepton numbers $L_\alpha$.
\item \refstepcounter{equation}
$\label{Z2tr}
\zz^{(\mu\tau)}: \ 
D_{\mu L} \leftrightarrow D_{\tau L}, \
\mu_R \leftrightarrow \tau_R, \
\nu_{\mu R} \leftrightarrow \nu_{\tau R}, \
\phi_3 \to - \phi_3$.
\hfill (\arabic{equation})
\item \refstepcounter{equation}
$\label{Z2aux}
\zz^{(\mathrm{aux})}: \
\nu_{eR}, \ \nu_{\mu R}, \ \nu_{\tau R}, \ \phi_1,\ e_R \ \mbox{change sign}.$
\hfill (\arabic{equation})
\end{enumerate}
In our notation,
$D_{\alpha L}$ denotes the left-handed-lepton gauge-$SU(2)$ doublets
and $\alpha_R$ denotes the right-handed charged-lepton $SU(2)$ singlets.
The symmetry $\zz^{(\mu\tau)}$ transposes the muon and tau family
and is spontaneously broken by the VEV of $\phi_3$;
as pointed out earlier,
that Higgs doublet is necessary in order to obtain $m_\mu \neq m_\tau$. 
The symmetry $\zz^{(\mathrm{aux})}$,
which is spontaneously broken by the VEV of $\phi_1$, 
is auxiliary and forbids the Yukawa coupling of $\phi_3$
to the $\nu_{\alpha R}$;
this is imperative for keeping the neutrino mass matrix
$\mu$--$\tau$-symmetric at tree level.

The above symmetries and multiplets
uniquely determine the Yukawa Lagrangian as 
\begin{equation}\label{L}
\begin{array}{rcl}
\mathcal{L}_\mathrm{Y} & = & 
- y_1 \bar D_{eL} \nu_{eR} \tilde\phi_1  
- y_2 \left( \bar D_{\mu L} \nu_{\mu R} + \bar D_{\tau L} \nu_{\tau R} \right)
\tilde\phi_1 
\\ && 
- y_3 \bar D_{eL} e_R \phi_1
- y_4 \left( \bar D_{\mu L} \mu_R + \bar D_{\tau L} \tau_R \right) \phi_2
- y_5 \left( \bar D_{\mu L} \mu_R - \bar D_{\tau L} \tau_R \right) \phi_3
+ \mbox{H.c.}
\end{array}
\end{equation}
Due to the symmetries $U(1)_{L_\alpha}$
the Yukawa couplings are flavour-diagonal.
There must be a source of non-trivial lepton mixing
and the obvious choice for this purpose
are the Majorana mass terms of the $\nu_{\alpha R}$.
These mass terms have dimension three,
therefore there will be \emph{soft}\ breaking
of the family lepton numbers.\footnote{The same mechanism was
used in ref.~\cite{GLsoft1} for the purpose of breaking $L_e - L_\mu - L_\tau$.} 
The mass Lagrangian of the $\nu_{\alpha R}$ is given by
\be\label{massL}
\mathcal{L}_\mathrm{Maj} = \frac{1}{2} \left( \begin{array}{ccc}
\nu_{eR}^T, & \nu_{\mu R}^T, & \nu_{\tau R}^T \end{array} \right) C^{-1} M_R^* 
\left( \begin{array}{c} \nu_{eR} \\ \nu_{\mu R} \\ \nu_{\tau R} 
\end{array} \right) + \mbox{H.c.}
\ee
In this model $M_R$ is the only source of lepton mixing.
Although $\mathcal{L}_\mathrm{Maj}$ breaks the family lepton numbers softly,
it has to preserve the $\mu$--$\tau$ symmetry.
Therefore,
the mass matrix $M_R$ must be $\mu$--$\tau$-symmetric. 
Consequently,
the only source of $\zz^{(\mu\tau)}$ breaking is the VEV of $\phi_3$. 

From equation~(\ref{L}) we read off that
the neutrino Dirac mass matrix
has the $\mu$--$\tau$-symmetric 
form $M_D = \mbox{diag} \left( c,\, d,\, d \right)$.
With the seesaw formula we obtain $\mathcal{M}_\nu$ as
\be
\mnu = -M_D^T M_R^{-1} M_D.
\ee
Since both $M_D$ and $M_R$ are $\mu$--$\tau$-symmetric,
the same applies to $\mnu$ which,
therefore,
has the form displayed in equation~(\ref{rtyde}).
Moreover,
since the charged-lepton mass matrix is diagonal,
the lepton mixing matrix is obtained
by diagonalizing $\mnu$ according to equation~(\ref{diag}). 

The symmetries on which the model is based generate
a non-Abelian symmetry group
because the $U(1)_{L_\alpha}$ with $\alpha = \mu, \tau$
do not commute with $\zz^{(\mu\tau)}$.
It is easy to show that
the non-Abelian core of the symmetry group is $O(2)$~\cite{GLsu5},
with $U(1)_{\left. \left( L_\mu - L_\tau \right) \right/ 2}$ and $\zz^{(\mu\tau)}$
corresponding,
respectively,
to the $U(1)$ and the reflection contained in $O(2)$.
In the present model,
the Higgs doublets transform
according to one-dimensional irreducible representations of $O(2)$.
For a viable model
with Higgs doublets in a two-dimensional irreducible representation of $O(2)$,
see ref.~\cite{GLNo2}.

An interesting property of the model in this section
is that lepton-flavour-changing processes
have \emph{finite}\ one-loop amplitudes~\cite{GLsoftLFV}.
The reason is that flavour violation
occurs only via \emph{soft}\ $L_\alpha$-breaking terms.

\section{A model based on a generalized CP transformation}

In this section we modify the model of the previous one
in order to obtain a model which reproduces the neutrino mass matrix
of equation~(\ref{rtydr}).
The new model has exactly the same multiplets
as the one of section~\ref{softly broken}.
The symmetries $U(1)_{L_\alpha}$ and $\zz^{(\mathrm{aux})}$ also apply.
The new feature is that $\zz^{(\mu\tau)}$ is replaced by the non-standard
(generalized)
CP transformation~\cite{HScp,GLcp} 
\begin{equation}\label{CP}
\begin{array}{rcl}
D_{\alpha L} &\to& i S_{\alpha \beta} \gamma^0 C \bar D_{\beta L}^T,
\\  
\nu_{\alpha R} &\to& i S_{\alpha \beta} \gamma^0 C \bar \nu_{\beta R}^T,
\\
\alpha_R &\to& i S_{\alpha \beta} \gamma^0 C \bar \beta_R^T,
\\ 
\phi_{1,2} &\to& \phi_{1,2}^\ast,
\\
\phi_3 &\to& - \phi_3^\ast,
\end{array}
\end{equation}
where $\alpha, \beta = e, \mu, \tau$
and $S$ is the matrix in equation~(\ref{S}).
The resulting Yukawa Lagrangian is very similar,
but not quite identical,
to the one of the previous model,
\textit{cf.}\ equation~(\ref{L}):
\ba
\mathcal{L}^\prime_\mathrm{Y} &=& 
- y_1 \bar D_e \nu_{eR} \tilde\phi_1  
- \left( y_2 \bar D_\mu \nu_{\mu R} + y_2^\ast \bar D_\tau \nu_{\tau R} \right)
\tilde\phi_1
\no & &
- y_3 \bar D_e e_R \phi_1
- \left( y_4 \bar D_\mu \mu_R +  y_4^\ast \bar D_\tau \tau_R \right) \phi_2
- \left( y_5 \bar D_\mu \mu_R -  y_5^\ast \bar D_\tau \tau_R \right) \phi_3 
+ \mbox{H.c.}
\label{Lcp}
\ea
The coupling constants $y_1$ and $y_3$ are real
whereas $y_2$,
$y_4$,
and $y_5$ are in general complex. 

The breaking of the symmetries proceeds in the same way
as in the previous model.
In particular,
the family lepton numbers are broken softly by
the Majorana mass terms of the $\nu_{\alpha R}$.
Because of the non-standard CP transformation of equation~(\ref{CP}),
the mass matrix $M_R$ in equation~(\ref{massL})
is of the same form of the $\mnu$ in equation~(\ref{rtydr}).
Moreover,
assuming,
without loss of generality,
that the VEV of $\phi_1$ is real,
we read off from the Lagrangian of equation~(\ref{Lcp}) that
$M_D = \mbox{diag} \left(c,\ d,\ d^\ast \right)$ with real $c$. 
Therefore,
$M_D$ fulfills $SM_DS = M_D^\ast$.
Since by construction the same is true for $M_R$,
the neutrino mass matrix resulting from the present model
is the one in equation~(\ref{rtydr}).

\section{On the small ratio of muon to tau mass}

In models with $\mu$--$\tau$ interchange symmetry
one would naively expect the muon and tau charged-lepton masses
to be of the same order of magnitude.
However,
this is not so,
since $m_\mu / m_\tau \simeq 1 / 18$.

Let us consider the muon and tau masses
in the models of refs.~\cite{GLmu-tau} and~\cite{GLcp}. 
Denoting the VEV of $\phi_j^0$ ($j=1,2,3$) by $v_j \left/ \sqrt{2} \right.$, 
the muon and tau masses are given by
\ba
m_\mu  &=& \frac{1}{\sqrt{2}} \left| y_4 v_2 + y_5 v_3 \right|,
\\
m_\tau &=& \frac{1}{\sqrt{2}} \left| y_4 v_2 - y_5 v_3 \right|,
\ea
respectively,
in the model of ref.~\cite{GLmu-tau},
\textit{cf.}\ equation~(\ref{L}),
while in the model of ref.~\cite{GLcp} those masses are given by
\ba
m_\mu &=& \frac{1}{\sqrt{2}} \left| y_4 v_2 + y_5 v_3 \right|,
\\
m_\tau &=& \frac{1}{\sqrt{2}} \left| y_4^\ast v_2 - y_5^\ast v_3 \right|.
\ea
Clearly,
one must use finetuning for obtaining $m_\mu \ll m_\tau$.
Moreover,
as one can read off from the formulas above, 
that finetuning is rather awkward
since one has to choose two products
of unrelated quantities---one Yukawa coupling
and one VEV---such that those two products nearly cancel in $m_\mu$.
In order to soften the amount of finetuning,
we have proposed in refs.~\cite{GLcp,GLsmallratio}
to add to the models the following symmetry: 
\be
K: \quad \mu_R \to -\mu_R, \ \phi_2 \leftrightarrow  \phi_3.
\label{K}
\ee
This symmetry leads to
\be
y_4 = - y_5
\label{yy}
\ee
in both the Lagrangians of equations~(\ref{L}) and~(\ref{Lcp}). 
In this way the finetuning is confined to the VEVs,
since
\be
\frac{m_\mu}{m_\tau} = \left| \frac{v_2 - v_3}{v_2 + v_3} \right|.
\label{ratio}
\ee
The symmetry $K$ has the nice property that
it can be directly implemented in both models;
its effect is simply to reduce the number of free parameters.

The symmetry $K$ also affects the scalar potential.
With $K$ the minimum of the potential features $v_2 = v_3$,
under certain conditions on the coupling constants~\cite{GLsmallratio}. 
Therefore,
$K$ leads to $m_\mu = 0$ and has to be broken.
The idea is to break it softly,
through terms of dimension two in the scalar potential.
With this mechanism, 
$m_\mu \ll m_\tau$ can be justified,
at least in a technically natural way,
since it will be the consequence of a small soft breaking.

There are some differences between the scalar potentials $V$ and $V^\prime$
of the models of refs.~\cite{GLmu-tau} and~\cite{GLcp},
respectively.
We first treat the simpler case of the model of ref.~\cite{GLmu-tau}.
The symmetries $\zz^{(\mathrm{tr})}$ in equation~(\ref{Z2tr})
and $\zz^{(\mathrm{aux})}$ in equation~(\ref{Z2aux})
require that in every term of $V$ each scalar doublet $\phi_j$
occurs an even number of times.
Requiring in addition invariance under the symmetry $K$ in equation~(\ref{K}),
one finds the potential
\ba
V_\phi
&=& - \mu_1 \phi_1^\dagger \phi_1
- \mu_2 \left( \phi_2^\dagger \phi_2 + \phi_3^\dagger \phi_3 \right)
\no & &
+ \lambda_1 \left( \phi_1^\dagger \phi_1 \right)^2
+ \lambda_2 \left[ \left( \phi_2^\dagger \phi_2 \right)^2
+ \left( \phi_3^\dagger \phi_3 \right)^2 \right]
\no & &
+ \lambda_3\, \phi_1^\dagger \phi_1
\left( \phi_2^\dagger \phi_2 + \phi_3^\dagger \phi_3 \right)
+ \lambda_4\, \phi_2^\dagger \phi_2\, \phi_3^\dagger \phi_3
\no & &
+ \lambda_5 \left(
\phi_1^\dagger \phi_2\, \phi_2^\dagger \phi_1
+ \phi_1^\dagger \phi_3\, \phi_3^\dagger \phi_1 \right)
+ \lambda_6\, \phi_2^\dagger \phi_3\, \phi_3^\dagger \phi_2
\no & &
+ \lambda_7 \left[ \left( \phi_2^\dagger \phi_3 \right)^2 + \mbox{H.c.} \right]
+ \left\{ \lambda_8 \left[ \left( \phi_1^\dagger \phi_2 \right)^2
+ \left( \phi_1^\dagger \phi_3 \right)^2 \right]
+ \mbox{H.c.} \right\}.
\label{Vphi}
\ea
All the coupling constants in $V_\phi$,
except for $\lambda_8$,
are real.
The term which conserves $\zz^{(\mathrm{tr})}$ and $\zz^{(\mathrm{aux})}$
but breaks $K$ softly is unique:
\be
V_\mathrm{soft} = \mu_\mathrm{soft}
\left( \phi_2^\dagger \phi_2 - \phi_3^\dagger \phi_3 \right).
\label{soft}
\ee
The full potential is $V = V_\phi + V_\mathrm{soft}$.
It can be shown~\cite{GLsmallratio} that
$\tilde\lambda \equiv - 2 \lambda_2 + \lambda_4 + \lambda_6 + 2 \lambda_7 < 0$
and $\lambda_7 < 0$ are sufficient conditions
for the minimum of $V_\phi$ to be at $v_2 = v_3$.
It is then straightforward to find the approximate relation
\be
\frac{m_\mu}{m_\tau} \simeq \left| \frac{2\mu_\mathrm{soft}}%
{\tilde \lambda \left( \left| v_2 \right|^2
+ \left| v_3 \right|^2 \right)} \right|,
\ee
which explicitly displays that a small $\mu_\mathrm{soft}$
leads to a small $m_\mu/m_\tau$.

Replacing $\zz^{(\mathrm{tr})}$
by the generalized CP transformation of equation~(\ref{CP})
allows two extra terms in the scalar potential:
\ba
V_9 &=& \lambda_9
\left( i \phi_1^\dagger \phi_2\, \phi_1^\dagger \phi_3 + \mathrm{H.c.} \right),
\label{V9} \\
V_{10} &=& \lambda_{10} 
\left( \phi_2^\dagger \phi_2 - \phi_3^\dagger \phi_3 \right)
\left( i \phi_2^\dagger \phi_3 + \mathrm{H.c.} \right),
\label{V10} 
\ea
with real $\lambda_9$ and $\lambda_{10}$.
Moreover,
in equation~(\ref{Vphi}) the coupling constant $\lambda_8$ must now be real.
To complicate matters,
besides the soft $K$-breaking term in equation~(\ref{soft}),
a second soft-breaking term is allowed,
which is given by
\be
V^\prime_\mathrm{soft} = \mu'_\mathrm{soft}
\left( i \phi_2^\dagger \phi_3 + \mathrm{H.c.} \right).
\label{soft'}
\ee
The full potential under discussion is then
$V^\prime = V_\phi + V_9 + V_{10} + V_\mathrm{soft} + V^\prime_\mathrm{soft}$.
Though it is now more involved,
it can still be demonstrated~\cite{GLsmallratio} that
a non-zero ratio $m_\mu/m_\tau$ arises at first order
in the soft $K$-breaking parameters $\mu_\mathrm{soft}$
and $\mu^\prime_\mathrm{soft}$.
One and the same mechanism is operative in both models.

\section{Summary}

In this review we have seen that a $\mu$--$\tau$-symmetric
effective light-neutrino Majorana mass matrix $\mnu$
leads to the predictions $\theta_{23} = \pi / 4$ and $U_{e3} = 0$.
Since the second prediction now seems to be in disagreement with phenomenology,
one may alternatively employ a $\mnu$ symmetric under a CP transformation 
involving the $\mu$--$\tau$ interchange symmetry;
in that case,
$\theta_{23} = \pi / 4$ and $\cos{\delta} = 0$ ensue.
One has in both cases,
 at the tree level,
maximal atmospheric neutrino mixing ($\theta_{23} = 45^\circ$),
a prediction which is in full agreement with experiment.

We have proceeded to show that both effective $\mnu$'s may be obtained
from simple models based on the type~I seesaw mechanism
and on family-lepton-number symmetries which are \emph{softly broken}\/
in the dimension-three Majorana mass matrix $M_R$
of the right-handed gauge-singlet neutrinos.
Those models may be furnished with an extra symmetry $K$,
whose \emph{soft breaking}\/ through dimension-two terms in the scalar potential
allows one to explain,
in a natural fashion,
the small but non-zero ratio of the muon and tau charged-lepton masses.

\begin{acknowledgement}
The work of L.L.\ is funded by the Portuguese
Foundation for Science and Technology (FCT)
through FCT unit 777 and through the project PTDC/FIS/
098188/2008. W.G.\ acknowledges support from the Austrian Science Fund
(FWF), Project Nr.\ P~24161-N16.
\end{acknowledgement}

\end{document}